\documentclass[twocolumn,showpacs,preprintnumbers,amsmath,amssymb]{revtex4}

\usepackage{epsfig,latexsym}

\newcommand{\sect}[1]{\setcounter{equation}{0}\section{#1}}

\begin{document}

\title{Toroidal Perturbations of Friedmann-Robertson-Walker Universes}

\author{Ji\v ri Bi\v c\'ak,$^{1,2,4}$
Donald Lynden-Bell,$^{2,1,4}$ and Joseph Katz$^{3,1,4}$
\\$^1${\it Institute of Theoretical Physics,  Charles University, \it V Hole\v sovick\' ach 2, 180 00 Prague 8, Czech Republic}
\\{\it $^2$Institute of Astronomy, Madingley Road, Cambridge CB3 0HA,
United Kingdom}
 \\ {\it $^3$The Racah Institute of Physics, Givat Ram, 91904 Jerusalem,
Israel}
\\ {\it $^4$Max-Planck Institute for Gravitational Physics (Albert Einstein
Institute) \it 14476 Golm, Germany.}}

\begin{abstract}

Explicit expressions are found for the axisymmetric metric perturbations  of the closed, flat and open FRW
universes caused by toroidal motions of the cosmic fluid. 
The perturbations are decomposed in vector spherical harmonics on 2-spheres,
but the radial dependence is left general. Solutions for
general odd-parity $l$-pole perturbations  are given for either angular velocities  or angular momenta 
prescribed. In particular, in case of closed universes the solutions require a
special treatment of the Legendre equation.

\end{abstract}

\maketitle

\sect{Introduction}

There exists a class of perturbations  in the Newtonian astrophysics which represent
differential rotations of a spherical star in hydrostatic equilibrium. The
displacement fields are time-independent and have no radial components
\cite{AS1}:
\begin{eqnarray}
4\pi G\rho\xi_r=0,\qquad 4\pi
G\rho\xi_\theta=\frac{T_{lm}(r)}{r\sin\theta}\frac{\partial Y_{lm}}{\partial\varphi},
\nonumber\\
\qquad 4\pi G\rho\xi_\varphi=-\frac{T_{lm}(r)}{r}\frac{\partial Y_{lm}}{\partial\theta},
\label{11}
\end{eqnarray}
where $T_{lm}$ is any function satisfying the condition that $\xi_\theta$ and
$\xi_\varphi$ are finite. These solutions define the {\it toroidal perturbations}.
When $l=1$, the displacements can be written as
\begin{equation}
{\bf \xi}=\delta{\bf\Omega}(r)\times {\bf
r},\qquad |\delta{\bf\Omega}|=\frac{1}{4\pi G\rho^2}T_{ml},
\end{equation}
which represent rigid rotations of the internal spherical surfaces around
the center. If $\delta{\bf\Omega}$ does not depend on $r$, the whole star is rotated as a rigid body.
To get a uniform slow rotation, rather than a static displacement,
we assume equation (\ref{11}) to be true with $\xi$ replaced by the Eulerian velocity
field; then displacements are linear in time. Such perturbations
are analogues of the cosmological toroidal perturbations investigated in this paper.

The toroidal perturbations  (\ref{11}) are transverse and divergence free. They are
often called the ``trivial modes". The trivial modes become important when one considers the
second order perturbations  ($\simeq[\delta{\bf\Omega}]^2$) \cite {S}. They develop
into the so-called $r$-modes in non-radial oscillations of a rotating star (see
e.g.\cite{PP}). They also give rise to non-zero eigenvalues when an external
(e.g. magnetic) field is added.

Toroidal oscillations are important in geophysics \cite{D}. Non-rotating,
spherically symmetric  Earth models admit ``trivial toroidal modes" which are
associated with vanishing eigenfrequencies; they do not alter the
elastic-gravitational potential of the Earth. As in astrophysics, their
counterparts play a significant role on a rotating Earth. In the solid core
the toroidal modes are nontrivial even in spherical models: the presence of
an anisotropic stress tensor produces a traction.

The velocity pattern of the axisymmetric  $l=2$ toroidal (quadrupole)
mode is shown in Fig. 1. The individual rings of fluid rotate independently, preserving
their local angular momentum. For non-trivial modes in the solid core the
displacements will reverse after a half of the oscillation cycle.

\begin{figure}[here]
\includegraphics{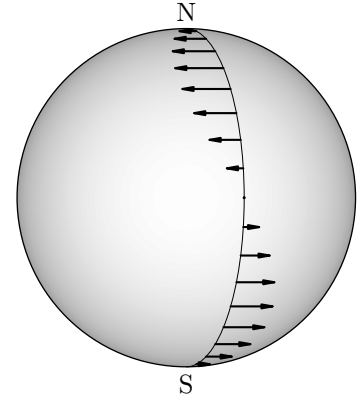}
\caption{Velocity field for axisymmetric quadrupole ($l=2$) toroidal perturbations.}
\end{figure}

In general relativity  even ``trivial modes" in spherical systems become ``non-trivial"
since they produce dragging of inertial frames and, hence, influence
essentially gravitational field.

In the following we investigate the toroidal perturbations  of Friedmann-Robertson-Walker (FRW)
universes of all three types $(k=-1,0,+1)$.  We confine ourselves to the
axisymmetric  case. Since the backgrounds  admit homogeneous  and isotropic foliations,
non-symmetric perturbations  can be found from the axisymmetric ones.

Indeed, in any spherically symmetric background system, one can restric oneself to
axisymmetric modes of perturbations since non-axisymmetric modes with an $e^{im\varphi}$
dependence ($m=-l,\ldots,+l$) can be derived from the axisymmetric perturbations ($m=0$) by
suitable rotation of the axes. If, in the original coordinates, the new polar axis will be
pointing in a direction ($\theta'$,$\varphi'$), then an axisymmetric mode with respect to
the new axes can be decomposed in non-axisymmetric modes in the original axes by using the
addition theorem for spherical harmonics (see \cite{Cha}, p. 138--139 for more detailes). Each 
the  $e^{im\varphi}$ component will separately solve the perturbation equation with the
original radial functions. Since we are studying vector perturbations, the vector
components will also transform under a rotation. (So, in (\ref{11}) with an 
axisymmetric mode only $\xi_\varphi\ne0$, but a general rotation will induce 
$\xi_\theta\ne0$ as well.)

The equations  to be solved are derived and discussed in Section 2 of \cite{BKL},
hereafter Paper~I. Here we just summarize few basic relations. We start out
  from the perturbed FRW metric in the form
\begin{eqnarray}
{\rm d}s^2&=&(\overline g_{\mu\nu}+h_{\mu\nu}){\rm d}x^\mu {\rm d}x^\nu
\\\nonumber
          &=& {\rm d}t^2-a^2(t)
f_{ij}{\rm d}x^i {\rm d}x^j+h_{\mu\nu}{\rm d}x^\mu {\rm d}x^\nu\;,
\end{eqnarray}
where the background  metric $\overline g_{\mu\nu}$ is used to move indices; the
time-independent part of the  spatial background metric
$f_{ij}~[i,j,k=1,2,3]$ is used to define the 3-covariant derivative $\nabla_k$,
$\nabla^k = f^{kl} \nabla_l$.   We choose spatial
harmonic gauge conditions on $t= \rm constant$ slices. The slices themselves are
chosen so that the perturbation  of the external curvature vanishes (uniform Hubble
expansion gauge) -- see Paper~I for details. In addition, since we are interested in the toroidal perturbations 
we also assume the perturbation  of the fluid velocity to be
transverse,$\nabla_kV^k=0$, and the same for the metric vector perturbation 
$h_{0k}$, $\nabla^kh_{0k}=0$.

In the axisymmetric  case we get the perturbed constraint equation 
\begin{equation}
\nabla^2h_{0\varphi}+2kh_{0\varphi}=2a^2\kappa\delta T^0_\varphi,
\label{12}
\end{equation}
in which $\nabla^2=f^{kl}\nabla_k\nabla_l$ and the perturbed Bianchi identities
(the dot denoting ${\partial}/{\partial t}$)
\begin{equation}
\left[   a^3(\rho+p)\left(   a^2r^2\sin^2\theta
V^\varphi-h_{0\varphi}\right)\right]^{\,\bf\dot{}}=0,
\end{equation}
equivalently
\begin{equation}
\left[   a^3\delta T^0_\varphi\right]^{\,\bf\dot{}}=0.
\label{14}
\end{equation}
These just express the conservation of angular momentum  of each element of the axially
symmetric  ring of fluid, analogously to the ``trivial modes" in astrophysics
mentioned above. To have solutions in more closed forms, we decompose
quantities only in vector spherical harmonics in variables $\theta,\varphi$, but
leave the radial dependence general. Hence, we write
\begin{eqnarray}
h_{0\varphi}&=&a^2r^2\sum\limits_{l=1}^{\infty}\omega_l(t,r)\sin\theta Y_
{l0,\theta},\label{17}
\\V_\varphi&=&-a^2r^2\sum\limits_{l=1}^{\infty}\Omega_l(t,r)\sin\theta
 Y_{l0,\theta},
\label{18}
\end{eqnarray}
and
\begin{eqnarray}
\delta
T^0_\varphi&=&a^2(\rho+p)r^2\sum\limits_{l=1}^{\infty}
(\omega_l-\Omega_l)\sin\theta~ Y_{l0,\theta}
\nonumber\\
&=&
\sum\limits_{l=1}^{\infty}[\delta
T^0_\varphi(t,r)]_l\sin\theta\/ Y_{l0,\theta}.
\label{19}
\end{eqnarray}
From equation  (\ref{12}) then follows the radial equation  for each $l$ in the form
\begin{eqnarray}
-\sqrt{1-kr^2}\frac{1}{r^2}\frac{\partial}{\partial r}\left[ 
\sqrt{1-kr^2}\frac{\partial}{\partial r}(r^2\omega_l)
\right]+
\nonumber\\
\frac{l(l+1)}{r^2}\omega_l-4k\omega_l=\lambda^2(\Omega_l-\omega_l),
\label{oml}
\end{eqnarray} 
where $\lambda^2=2\kappa a^2 (\rho+p)=4(k-a^2 \dot H)$, 
$H = \dot a/ a$ is the Hubble constant; the last relation 
follows from the background  equations  for arbitrary $\rho$, $p$, $k$ and
cosmological constant $\Lambda$.

For $k=0$ equation (\ref{oml}) can be written with the
angular momentum  density $(\delta T^0_\varphi)_l$ as a source in the form 
\begin{eqnarray}
\frac{1}{r^4}\frac{\partial}{\partial r}\left(  r^4 \frac{\partial\omega_l}{\partial
r}\right)-\frac{l(l+1)-2}{r^2}\omega_l&=&\lambda^2(\omega_l-\Omega_l)
\hphantom{AAAA}
\nonumber\\
&=&\frac{2\kappa}{r^2}(\delta
T^0_\varphi)_l\;,
\label{oml8}
\end{eqnarray}
for the fluid angular velocity as a source,
\begin{eqnarray}
\frac{1}{r^4}\frac{\partial}{\partial r}\left(  r^4 \frac{\partial\omega_l}{\partial
r}
\right)-\left[\lambda^2+\frac{l(l+1)-2}{r^2}\right]
\omega_l&&
\nonumber\\
=-\lambda^2\Omega_l\; ;&&
\label{oml2}
\end{eqnarray}
here $\lambda^2=-4a^2\dot H=2\kappa a^2(\rho+p)$. 

For $k=\pm 1$ it is advantageous to write
$r^2=k(1-\mu^2)$, or $\mu=\sqrt{1-kr^2}$, to obtain
\begin{eqnarray}
\frac{1}{[k(1-\mu^2)]^{3/2}}\frac{\partial}{\partial\mu}\Big\{ 
[k(1-\mu^2)]^{5/2}\frac{\partial\omega_l}{\partial\mu}\Big\}
-\frac{l(l+1)-2}{k(1-\mu^2)}\omega_l&\!\!&\!\!
\nonumber\\
=\frac{2\kappa}{k(1-\mu^2)}(\delta
T^0_\varphi)_l.\hphantom{AAAAA}&\!\!&\!\!
\label{oml3}
\end{eqnarray}
If we set
\begin{equation}
\omega_l=[k(1-\mu^2)]^{-3/4}\overline \omega_l\;,
\label{241}
\end{equation}
 equation  (\ref{oml3}) becomes the Legendre equation 
 for $\bar \omega_l$ with $(\delta T^0_\varphi)_l$ as the source:
\begin{eqnarray}
&&\frac{\partial}{\partial\mu}\left[
k(1-\mu^2)\frac{\partial\overline\omega_l}{\partial\mu}\right]+\left[ k \frac{3}{2}
(\frac{3}{2}+1)-
\frac{(l+\frac{1}{2})^2}{k(1-\mu^2)}
\right] \overline\omega_l
\nonumber\\
&&\hphantom{AAAAAAAAAAAA}
=\frac{2\kappa}{[k(1-\mu^2)]^{1/4}}(\delta T^0_\varphi)_l.
\label{242}
\end{eqnarray}
With the fluid angular velocity $\Omega_l$ as the source, the equation reads
\begin{eqnarray}
&&\frac{\partial}{\partial\mu}\left[ 
k(1-\mu^2)\frac{\partial\overline\omega_l}{\partial\mu}\right]+\left[ k\nu(\nu+1)-
\frac{(l+\frac{1}{2})^2}{k(1-\mu^2)}
\right]\overline\omega_l
\nonumber\\
&&\hphantom{AAAA}
=-K_l\equiv-\lambda^2\Omega_l[k(1-\mu^2)]^{3/4},
\label{oml4}
\end{eqnarray}
where
\begin{equation}
\left(\nu+\frac{1}{2}\right)^2=4-2k\kappa a^2(\rho+p)=4-k\lambda^2=4ka^2\dot H.
\label{244}
\end{equation}

In the following we shall first solve equations (\ref{oml8}) and  (\ref{oml2})
for flat universes (Section 2). The interesting but more complicated case is that of closed universes, 
i.e., solutions of equations  (\ref{242})  and (\ref{oml4}) with $k=+1$.
We shall discover that Legendre functions $Q^{l+\frac{1}{2}}_{\nu=3/2}$
which solve equation  (\ref{242}) and are regular at least at one pole on the 3-sphere,
in fact vanish identically for $l\ge 2$!
However, appropriate solutions can be found by considering the derivatives of the Legendre functions
with respect to their degree $\nu$. We did not find this fact in the literature on special
functions. We analyze solutions of both equation  (\ref{242}) with the angular momentum given and
(\ref{oml4}) with the fluid angular velocity given for closed universes in Section 3.
Finally, in Section 4, we study them in detail also for open universes.
Roughly speaking, the dragging effects are not damped when angular momenta 
are given as their sources, whereas they are, if fluid's angular velocity is considered as a source.
The physical explanation of this apparent paradox is the same as that given in Paper~I for $l=1$
perturbations.

Before we turn to solving the equations, we wish to emphasize that all our solutions are true at any
given instant of time. However, the angular momentum and angular velocity distributions
are related to those at earlier times by equations of motion (i.e., by the contracted Bianchi
identities). In our case of toroidal perturbations they lead back to local conservation
of angular momentum density, which demonstrates the advantage of taking angular momentum as the source.
The time evolution of the dragging is thus independent of $l$. It is analyzed in Section 5 of Paper~I.


\sect{Axisymmetric toroidal perturbations for $k=0$}

\renewcommand{\thesubsection}{\Alph{subsection}}
\subsection{The angular momentum  as a source of $\omega$}

The homogeneous equation  corresponding to (\ref{oml8}) with $(\delta
T^0_\varphi)_l$ given is
\begin{equation}
\omega_l''+\frac{4}{r}\omega_l'-\frac{l(l+1)-2}{r^2}\omega_l=0.
\end{equation}
This can be easily solved by writing $\omega_l=r^\alpha\overline\omega_l$, and choosing
$\alpha$ so that the coefficient of $\overline\omega_l$ vanishes. It implies $\alpha=l-1$
or $\alpha=-(l+2)$ and the first-order equation  for $\overline\omega_l'$. Two simple
integrations lead to the solutions
\begin{equation}
\omega^{(I)}_l=A_lr^{l-1}, \qquad \omega^{(II)}_l=B_lr^{-(l+2)},
\end{equation}
$A_l,B_l$ constants. The Wronskian is ${\cal W}_l=A_lB_l(2l+1)r^{-4}$. The
solution of the inhomogeneous equation  (\ref{oml8}) which decays at infinity
and is well-behaved at the origin is given by
\begin{eqnarray}
\omega_l=\frac{-2\kappa}{2l+1}\left[   \frac{1}{r^{l+2}}\int_0^r (r'){^{l+1}}(\delta
T^0_\varphi)_l \,{\rm d}r' \right.
\nonumber\\
+ \left. r^{l-1}\int_r^\infty (r')^{-l}(\delta
T^0_\varphi)_l \,{\rm d}r'\right].
\label{63}
\end{eqnarray}
In particular for the dipole $(l=1)$ perturbations 
\begin{equation}
\omega_{l=1}=-\frac{16\pi}{3}\left[   \frac{1}{r^3}\int_0^r\!\! (r')^2(\delta
T^0_\varphi)_1\,{\rm d}r' + \int_r^\infty\!\! \frac{1}{r'}(\delta
T^0_\varphi)_1\,{\rm d}r' \right].\label{omegal1}
\end{equation}
If we write
\begin{equation}\delta T^0_\varphi=(\delta
T^0_\varphi)_1\sin\theta (-\sqrt{3/4\pi}\sin\theta),
\end{equation}
and integrate over $\theta$, we find that (\ref{omegal1}) coincides with the second equation in (I.3.4);
herafter, we denote in this way equations in Paper~I.

Equation  (\ref{63}) demonstrates how for general $l$-pole
distributions, ``local" angular momenta, determined by $(\delta
T^0_\varphi)_l$, at all distances contribute instantaneously to the dragging of
inertial frames  without any exponential cut-off or influence of any horizon --
see the detailed discussion of this point for $l=1$ perturbations in Paper~I.
Notice that angular momenta $(\delta T^0_\varphi)_l$ for $l>1$ do not, however, contribute to the {\it
total} angular momentum  in a spherical layer $\left<r_1,r_2\right>$ given by 
[cf. (I.2.24) and (I.2.25)]
\begin{equation}
{J}(r_1,r_2)=-2\pi\int_{r_1}^{r_2}{\rm d}r\int_0^\pi
{\rm d}\theta ~a^3r^2\sin\theta\delta T^0_\varphi.
\label{66}
\end{equation}
Substituting for $\delta T^0_\varphi$
from equation (\ref{19}), integrating by parts over $\theta$ and
realizing that
\begin{equation}
\int_0^\pi Y_{l0}\cos\theta\sin\theta \,{\rm d}\theta=\sqrt{4\pi/3}\;\delta_{l1}
\end{equation}
due to the orthogonality of spherical harmonics, expression (\ref{66}) reduces to
\begin{eqnarray}
{J}(r_1,r_2)&=&4\sqrt{\pi/3}~a^3\int_{r_1}^{r_2}{\rm d}r r^2(\delta T^0_\varphi)_{l=1}
\\\nonumber
&=&
4\sqrt{\pi/3}~a^5(\rho+p)\int_{r_1}^{r_2}{\rm d}r r^4(\omega_{l=1}-\Omega_{l=1}).
\label{67}
\end{eqnarray}
Such result is understandable on intuitive grounds: e.g for $l=2$ odd-parity
perturbations  rotations at antipodal $\theta$'s are equal and opposite --
the ``northern" hemisphere rotates in the opposite direction to the ``southern" hemisphere (see Fig.~1).

\subsection{The fluid angular velocity  as a source}

The equation  to be solved is (\ref{oml2}). We introduce
$z=\lambda r$ and assume $\lambda>0$, so $z\ge0$. Putting then
\begin{equation}
\omega_l=z^{-3/2}\overline\omega_l,
\label{68}
\end{equation}
the homogeneous part of  (\ref{oml2}) turns into the equation 
\begin{equation}
\frac{{\rm d}^2\overline\omega_l}{{\rm d}z^2}+\frac{1}{z}\frac{{\rm d}\overline\omega_l}{{\rm d}z}-\left[
1+\frac{(l+\frac{1}{2})^2}{z^2}\right] \overline\omega_l=0.
\end{equation}
This is the modified Bessel equation, the general solution being given by
\begin{equation}
\overline\omega_l=C_l I_{l+\frac{1}{2}}(z)+D_l K_{l+\frac{1}{2}}(z),
\label{610}
\end{equation}
where $ I_{l+\frac{1}{2}}$ and $ K_{l+\frac{1}{2}}$ are modified Bessel functions. For
an integer $l$ these functions  are given explicitly by finite sums (see e.g. 8.467,
8.468 in \cite{GR}, for still shorter expressions in terms of $(d/dz)^l$, see
\cite{MO}, \cite{BE}) which imply the following asymptotic expansions at $z\rightarrow\infty$,
\begin{eqnarray}
I_{l+\frac{1}{2}}(z)&=&\frac{e^z}{\sqrt{2\pi z}}+\cdots,
\nonumber\\
K_{l+\frac{1}{2}}(z)&=&\sqrt{\frac{\pi}{2z}}e^{-z}+\cdots,
\label{611}
\end{eqnarray}
so, independent of $l$ in the leading terms at infinity.
At the origin $z\rightarrow 0$,
\begin{eqnarray}
I_{l+\frac{1}{2}}(z)&=&\sqrt{\frac{2}{\pi}}\frac{1}{(2l+1)!!}z^{l+\frac{1}{2}}\!\left[ 1
+\frac{1}{2(2l+3)}z^2+\cdot\cdot\cdot\right],
\nonumber\\
K_{l+\frac{1}{2}}(z)&=&\sqrt{\frac{\pi}{2}}\frac{(2l)!}{2^ll!}\frac{1}{z^{l+\frac{1}{2}}}+\cdot\cdot\cdot\;.
\label{612}
\end{eqnarray}

The general solution of the original homogeneous equation  is thus given by
$\omega_l=\omega_l^{(I)}+\omega_l^{(II)}$, where
\begin{eqnarray}
\omega_l^{(I)}&=&A_l\overline
I_{l+\frac{1}{2}}(z)=A_lz^{-3/2}I_{l+\frac{1}{2}}(z),\label{613}\\
\omega_l^{(II)}&=&B_l\overline K_{l+\frac{1}{2}}(z)=B_lz^{-3/2}K_{l+\frac{1}{2}}(z),
\label{614}
\end{eqnarray}
($\overline I_{l+\frac{1}{2}}$ and $\overline K_{l+\frac{1}{2}}$ are defined by these relations).
$\omega_l^{(I)}$ behaves well at $z\rightarrow 0$, $\omega_l^{(II)}$ at $z\rightarrow\infty$.
It is easy to see that for $l=1$ one obtains the asymptotic expansions given
in Paper~I. The Wronskian reads (for the Wronskian of 
$I_{l+\frac{1}{2}}$ and $ K_{l+\frac{1}{2}}$ see \cite{MO}, p. 68)
\begin{equation}
{\cal W}_l(z)= -A_lB_lz^{-4}.
\label{615}
\end{equation}

Using again the method of variation of parameters to find the solution of
the inhomogeneous equation  (\ref{oml2}) which is well behaved at $z\rightarrow 0$
and vanishes at $z\rightarrow\infty$, we arrive at
\begin{eqnarray}
\omega_l&=&\overline K_{l+\frac{1}{2}}(z)\int_0^z(z')^4\overline I_{l+\frac{1}{2}}(z')\Omega_l(z')\,{\rm d}z'
\nonumber\\
&+&\overline I_{l+\frac{1}{2}}(z)\int_z^\infty (z')^4\overline K_{l+\frac{1}{2}}(z')\Omega_l(z')\,{\rm d}z',
\label{616}
\end{eqnarray}
$\overline I_{l+\frac{1}{2}}$ and $\overline K_{l+\frac{1}{2}}$ are defined by (\ref{613}) and
(\ref{614}).

The solution (\ref{616}) for
general $l$ can be explored in a way similar to that for $l=1$, as was done
in Paper~I. So, to find $\omega$
 for small $z$ with sources $\Omega_l$ not close, we neglect the first term in
(\ref{616}) and expand $\overline I_{l+\frac{1}{2}}$ in the second term:
\begin{eqnarray}
\omega_l(z)&=&\sqrt{\frac{2}{\pi}}\frac{1}{(2l+1)!!}z^{l-1}\left[ 1
+\frac{1}{2(2l+3)}z^2+\cdots\right]\times
\nonumber\\
&\times&\int_z^\infty (z')^4\overline
K_{l+\frac{1}{2}}(z')\Omega_l(z')\,{\rm d}z'.
\label{617}
\end{eqnarray}
For the source $\Omega_l$ at $z'\gg1$ we expand $\overline K_{l+\frac{1}{2}}$ [see
(\ref{611})]:
\begin{eqnarray}
\omega_l(z)&=&\frac{1}{(2l+1)!!}z^{l-1}\left[ 1
+\frac{1}{2(2l+3)}z^2+\cdots\right]\times
\nonumber\\
&\times&\int_z^\infty
(z')^2e^{-z'}\Omega_l(z')\,{\rm d}z'.
\label{618}
\end{eqnarray}
If, in addition, the source is localized in a small interval $r_0(1\pm\Delta)$,
$\Delta\ll 1/\lambda$, we find
\begin{eqnarray}
\omega_l(r)&=&\frac{1}{(2l+1)!!}(\lambda r)^{l-1}\left[ 1
+\frac{1}{2(2l+3)}(\lambda r)^2\right]\times
\nonumber\\
&\times&(\lambda r_0)^3e^{-\lambda r_0}\; \overline\Omega\; 2\Delta.
\label{619}
\end{eqnarray}
This generalizes equation (I.4.7) for any $l$. Thus, we recover the
exponential decay which is now seen to be independent of $l$, but we also
discover the power-law decline, given by $(\lambda
r)^{l-1}/(2l+1)!!$ ($\lambda r\ll1$) which increases with increasing $l$. When
$\Omega_l$  is concentrated near $z_0$, 
in $z_0\pm\lambda\Delta$, then close to $z_0\gg1$ and with
$\Omega_l=\overline\Omega_l$ we find $\omega_l(z_0)=\lambda\Delta\overline\Omega_l$, 
as in Paper~I.


\sect{Axisymmetric toroidal perturbations for $k=1$}

We wish to solve the inhomogeneous equation  (\ref{oml3}) for $\omega_l$ which,
after the  substitution  (\ref{241}), turns into the inhomogeneous equation 
corresponding either to (\ref{242}) or (\ref{oml4}).

So, to find solutions we have first to explore appropriate solutions of the
homogeneous Legendre equation 
\begin{equation}
\frac{\partial}{\partial\mu}\left[   (1-\mu^2)\frac{\partial\overline\omega_l}{\partial\mu}\right]+\left[
\nu(\nu+1)-\frac{(l+\frac{1}{2})^2}{1-\mu^2}\right] \overline\omega_l=0
\label{620}
\end{equation}
for $l=1,2,3,\cdots$ and either $\nu=\frac{3}{2}$, or a general $\nu$
given by (\ref{244}).
Since for the closed universe the variable $\mu=\cos\chi$ is real and
$\mu\in \left< -1,+1 \right>$, we are interested in Legendre functions
$P^{l+\frac{1}{2}}_\nu(\mu)$, $Q^{l+\frac{1}{2}}_\nu(\mu)$ on the cut.
(We emphasize here that the order $\mu$ of the Legendre function, e.g. in $Q^\mu_\nu$,
has nothing to do with the independent variable $\mu$.)
Many useful formulas are given in \cite{MO}. In particular, we start from the general
expression for $Q^\mu_\nu$ in terms of a particular combination  of two
hypergeometric functions (given as the 4th expression on p. 168 in
\cite{MO}), change $\theta\rightarrow\chi$ and put $\mu=l+\frac{1}{2}$. Employing then
the well-known properties of hypergeometric functions  for special values of
parameters, we find, after some algebra, that functions 
$Q^{l+\frac{1}{2}}_\nu(\cos\chi)$ can be written as finite sums of the form
\begin{widetext}
\begin{eqnarray}
Q^{l+\frac{1}{2}}_\nu(\cos\chi)&=&\!\left(\frac{\pi}{2}\right)^{\frac{1}{2}}\!\frac{1}{\sin^{1/2}\chi}
\frac{\Gamma(\frac{3}{2}+\nu+l)}{\Gamma(\frac{3}{2}+\nu)}
\sum\limits _{k=0}^{l}
\frac{(-l)_k(l+1)_k}{(\frac{3}{2}+\nu)_k k!}\left(
\frac{\llap{$-$}1}{2\sin\chi}\right)^k
\!\!\cos\left[
(k+\frac{1}{2}+\nu)\chi+(k+l+1)\frac{\pi}{2}\right],
\label{621}
\end{eqnarray}
\end{widetext}
where the standard notation
$(\alpha)_k=\alpha(\alpha+1)\cdots(\alpha+k-1)=\Gamma(\alpha+k)/\Gamma(\alpha)$ is used. In
\cite{MO}, p.~168, suitable expression in terms of hypergeometric functions is
given also for Legendre functions  $P^\mu_\nu$ on the cut which one may wish to
use as independent solutions.  However, after putting $\mu=l+\frac{1}{2}$ and
making arrangements similar to those above for $Q^{l+\frac{1}{2}}_\nu$, it turns
out that $P^{l+\frac{1}{2}}_\nu$ diverge at $\chi=0,\pi$. [For $l=0,1$ this was
shown in Paper~I.] Therefore, only functions  $Q^{l+\frac{1}{2}}_\nu$
given by (\ref{621}) will be appropriate objects in the following.

It is easy to check that (\ref{621}) for $l=0$ implies equation (I.4.9). A small
calculation shows that  (\ref{621}) gives
\begin{eqnarray}
Q^{3/2}_{n-\frac{1}{2}}&=&\left( \frac{\pi}{2} \right)^\frac{1}{2}\frac{1}{\sin^{3/2}\chi}
\times
\label{622}
\\\nonumber
&\times&{\Big\{}-(n+1)\sin\chi\cos(n\chi)+\sin\left[ (n+1)\chi \right]{\Big\}},
\end{eqnarray}
which, after simple rearrangements, coincides with the first expression in
equation  (I.4.12) for $l=1$. As in Paper~I, we have put $\nu=n-\frac{1}{2}$ in
(\ref{622}) and will do so hereafter.

With $(\delta T^0_\varphi)_l$ given and $k=+1$, the homogeneous equation 
corresponding to (\ref{242}) becomes (\ref{620}) with $\nu=\frac{3}{2}$,
$n=2$. We thus will consider solutions $Q^{l+\frac{1}{2}}_{n-\frac{1}{2}}$, $n=2$,
determined by (\ref{621}). In Paper~I these solutions have been used for
$l=1$ and general $n$
 to determine $\omega_{l=1}$ when $\Omega_{l=1}$ is considered as a source.
Since we also wish to consider solutions $\omega_l$ for $\Omega_l$ as a source
with any $l$, we do not put $n=2$
yet, rather give first explicitly few functions 
\begin{equation}
{\cal Q}^l_n \stackrel{\rm def}= Q^{l+\frac{1}{2}}_{n-\frac{1}{2}}
\end{equation}
for general $n$ and $l=1,2,\cdots$, using the expression (\ref{621}). As with $l=1$,
we shall also consider $n$ purely imaginary, $n=iN$. Calculations
become lengthy with $l$ increasing. We have used MATHEMATICA for
checking, simplifying and deriving some of the following
formulas.

Starting first from (\ref{621}) for $l=1$, we get
\begin{widetext}
\begin{eqnarray}
{\cal Q}^1_n&=&\hphantom{i}\left( \frac{\pi}{2} \right)^\frac{1}{2}\frac{1}{\sin^{3/2}\chi}
\left[\cos\chi\sin(n\chi)-n\sin\chi\cos (n\chi) \right],
\label{624}\\
{\cal Q}^1_{iN}&=&i\left( \frac{\pi}{2} \right)^\frac{1}{2}\frac{1}{\sin^{3/2}\chi}
\left[\cos\chi\sinh(N\chi)-N\sin\chi\cosh (N\chi) \right].
\label{625}
\end{eqnarray}
For $l=2$ we obtain
\begin{eqnarray}
{\cal Q}^2_n&=&\hphantom{i}\left( \frac{\pi}{2} \right)^\frac{1}{2}\frac{1}{\sin^{5/2}\chi}
\left\{-\frac{3}{2}n\sin(2\chi)\cos(n\chi)+[2-n^2
\sin^2\chi+\cos(2\chi)]\sin (n\chi)
\right\},
\label{626}
\\
{\cal Q}^2_{iN}&=&i\left( \frac{\pi}{2} \right)^\frac{1}{2}\frac{1}{\sin^{5/2}\chi}
\left\{-\frac{3}{2}N\sin(2\chi)\cosh(N\chi)+[2+N^2
\sin^2\chi+\cos(2\chi)]\sinh (N\chi)\right\}.
\label{627}
\end{eqnarray}
With $l=3$ in (\ref{621}) lengthier calculations lead to
\begin{eqnarray}
{\cal Q}^3_n&=&\left( \frac{\pi}{2} \right)^\frac{1}{2}\frac{1}{\sin^{7/2}\chi}
\,\frac{1}{2}\Big\{n\left[(-n^2-11)\cos(2\chi)+n^2-19\right]
\cos(n\chi)\sin\chi
\nonumber
\\
&&\qquad +6\left[
(n^2+1)\cos(2\chi)-n^2+4\right]\cos\chi\sin (n\chi)\Big\},
\label{628}\\
{\cal Q}^3_{iN}&=&i\left( \frac{\pi}{2} \right)^\frac{1}{2}\frac{1}{\sin^{7/2}\chi}
\,\frac{1}{2}\Big\{N\left[(N^2-11)\cos(2\chi)-N^2-19\right]
\cosh(N\chi)\sin\chi\nonumber\\
&&\qquad +6\left[
(-N^2+1)\cos(2\chi)+N^2+4\right]\cos\chi\sinh (N\chi)\Big\}.
\label {629}
\end{eqnarray}
Finally, explicit expressions implied by (\ref{621}) for $l=4$ are
\begin{eqnarray}
{\cal Q}^4_n&=&\left( \frac{\pi}{2} \right)^\frac{1}{2}\frac{1}{\sin^{9/2}\chi}
\,\frac{1}{8}\Bigg\{20n\left[(n^2+5)\cos(2\chi)-n^2+16\right]
\cos(n\chi)\sin(2\chi) \nonumber\\
&+&\Bigl[ -3(n^4-25n^2+144)+
4(n^4-10n^2-96)\cos(2\chi)
-(n^4+35n^2+24)\cos(4\chi)\Bigr]\sin(n\chi)\Bigg\},
\label{630}
\\
{\cal Q}^4_{iN}&=&i\left( \frac{\pi}{2} \right)^\frac{1}{2}\frac{1}{\sin^{9/2}\chi}
\,\frac{1}{8}\Bigg\{20N\left[(-N^2+5)\cos(2\chi)+N^2+16\right]
\cosh(N\chi)\sin(2\chi) \nonumber\\
&+&\Bigl[-3(N^4+25N^2+144)+
4(N^4+10N^2-96)\cos(2\chi)
-(N^4-35N^2+24)
\cos(4\chi)\Bigr]\sinh(N\chi)\Bigg\}.
\label{631}
\end{eqnarray}
\end{widetext}
The behavior of these solutions at the poles $\chi=0,\pi$, as well as the
second independent solutions to the homogeneous Legendre equation 
(\ref{620}), will be discussed below.

Now we have to distinguish the case when $ (\delta T^0_\varphi)_l$ or $\Omega_l$
is considered as a source.
\subsection{The angular momentum  as a source for $\omega$}

In order to construct solutions to (\ref{242}) for $k=+1$, we need solutions
of the homogeneous Legendre equation  (\ref{620}) for $l=1,2,3,\ldots$ and
$\nu=\frac{3}{2}$, $n=2$, i.e. ${\cal Q}^l_2=Q^{l+\frac{1}{2}}_{\frac{3}{2}}$. For $l=1$
we have the solution  (\ref{622}) with $n=2$:
\begin{equation}
{\cal Q}^1_2=Q^{\frac{3}{2}}_{\frac{3}{2}} = 2\left( \frac{\pi}{2} \right)^\frac{1}{2}\sin^\frac{3}{2}\chi.
\label{632}
\end{equation}
This is well-behaved at both $\chi=0$ and $\chi=\pi$, whereas the second
(independent) solution
\begin{equation}
P^{\frac{3}{2}}_{\frac{3}{2}}
=-\left( \frac{\pi}{2} \right)^\frac{1}{2}\frac{\cos\chi}{\sin^\frac{3}{2}\chi}(1+2\sin^2\chi)
\end{equation}
diverges at these poles. The solution ${\cal Q}^1_2$ implies [cf. equation 
(\ref{241})] $\omega_{l=1}=\omega_0=$ constant, independent of $\chi$, as the
solution of the original equation  associated with (\ref{oml3}). Such $\omega_0$
may be time-dependent and may always enter $\omega$ in the case of closed
universes, in which an ``absolute rotation'' does not occur. It cannot be removed (see
equation  (I.3.8), and text below). The
other homogeneous solution for $\omega_1$, corresponding to
$P^{3/2}_{3/2}$, diverges at the poles. For $l=1$ the inhomogeneous equation  (\ref{242}) can then
be integrated directly. The integration yields
\begin{equation}
\frac{d\omega_1}{d\chi}=\frac{2\kappa}{\sin^4\chi}\int_0^\chi
\sin^2\chi'(\delta T^0_\varphi)_1\,{\rm d}\chi'.
\label{633}
\end{equation}
Starting from the definition of the angular momentum  ${J}(\chi_1,\chi_2)$ in
(I.2.25), with $r=\sin\chi$, and integrating over $\theta$ we get
\begin{equation}
{J}(0,\chi)= \frac{8\pi}{3}\sqrt{\frac{3}{4\pi}}a^3\int_0^\chi\sin^2\chi'(\delta
T^0_\varphi)_1\,{\rm d}\chi'.
\label {634}
\end{equation}
Since (as before) $\omega_{l=1}=-\sqrt{\frac{4\pi}{3}}\omega$ and $\kappa=8\pi$,
equation  (\ref{633}) becomes the equation (I.3.6), and we thus recover 
the solution (I.3.8).

Turning now to $l\ge2$ we are confronted with an intriguing problem.
Direct calculations (checked and for $l=4$ done by MATHEMATICA) show
that for $n=2$ all functions  ${\cal Q}^2_2$, ${\cal Q}^3_2$, ${\cal Q}^4_2$ given by
(\ref{626}), (\ref{628}) and (\ref{630}) vanish. In other words, Legendre functions 
$Q^{5/2}_{3/2}$, $Q^{7/2}_{3/2}$, $Q^{9/2}_{3/2}$ do vanish on the cut!
Using the recurrence relation (see \cite{MO}, p.171, with $P^\mu_\nu\rightarrow
Q^\mu_\nu$, and here we write $x$ instead of $\mu=\cos\chi$)
\begin{eqnarray}
Q^{\mu+2}_\nu(x)&+&2(\mu+1)\frac{x}{(1-x^2)^\frac{1}{2}}Q^{\mu+1}_\nu(x)
\nonumber\\&+&
(\nu-\mu)(\nu+\mu+1)Q^\mu_\nu(x)=0,
\label{635}
\end{eqnarray}
we can prove that in fact all
\begin{equation}
Q^{l+\frac{1}{2}}_{3/2}(x)=0~~~{\rm for}~~~l\ge 2.
\label{636}
\end{equation}
We are not aware of a mathematical reference where this is noticed.

How then to find regular solutions of the Legendre equation  (\ref{620}) for
$\nu=\frac{3}{2}$, $n=2$ and $l\ge2$? We start from the following
observations: if $Q^\mu_\nu$ solves Legendre equation  (\ref{620}) but for
some special values $\nu=\nu_0$ this $Q^\mu_\nu$ vanishes, then the
function $(\partial Q^\mu_\nu/\partial\nu)_{\nu=\nu_0}$ solves the equation  for
$\nu=\nu_0$ and may be non-vanishing. The proof can be done easily by a
direct calculation. Therefore, we wish to calculate $\partial{\cal Q}_n^l/\partial n$
from the expressions (\ref{626}), (\ref{628}), (\ref{630}), and put $n=2$.
Denoting
\begin{equation}
\tilde{\cal Q}^l_2=\frac{\partial{\cal Q}^l_n}{\partial n}\Bigg|_{n=2}=\frac{\partial
Q^{l+\frac{1}{2}}_{n-\frac{1}{2}}}{\partial n}\Bigg|_{n=2},
\label {637}
\end{equation}
we arrive at the following expressions:
\begin{equation}
\tilde{\cal Q}^2_2=\left( \frac{\pi}{2} \right)^\frac{1}{2}\frac{1}{\sin^{5/2}\chi}
\Big[-\frac{1}{4}\sin(4\chi)+2\sin(2\chi)-3\chi\Big],\label{638}
\end{equation}
\begin{eqnarray}
\tilde{\cal Q}^3_2&=&\left( \frac{\pi}{2} \right)^\frac{1}{2}\frac{1}{\sin^{7/2}\chi}\times
\hphantom{UUUUUUUUUUUUUUU}
\label {639}
\\
\nonumber
&\times&\Big[
\frac{1}{8}\sin(5\chi)-\frac{15}{8}\sin(3\chi)-10\sin\chi+15\chi\cos\chi\Big],
\end{eqnarray}
\begin{eqnarray}
\tilde{\cal Q}^4_2&=&\left( \frac{\pi}{2} \right)^\frac{1}{2}\frac{1}{\sin^{9/2}\chi}
\Big[-\frac{1}{8}\sin(6\chi)+3\sin(4\chi)+
\hphantom{AAAA}
\nonumber\\
&+&\frac{375}{8}\sin(2\chi)-15\chi[3\cos(2\chi)+4]\Big].
\label{640}
\end{eqnarray}
MATHEMATICA yields nice expansions at $\chi\rightarrow0$:
\begin{eqnarray}
\tilde{\cal Q}^2_2&=&\left( \frac{\pi}{2} \right)^\frac{1}{2}\left[
-\frac{8}{5}\chi^{5/2}+\frac{2}{21}\chi^{9/2}+{\cal
O}(\chi^{11/2})\right],\hphantom{AAAA}\label{641}
\\
\tilde{\cal Q}^3_2&=&\left( \frac{\pi}{2} \right)^\frac{1}{2}\left[
-\frac{8}{7}\chi^{7/2}-\frac{2}{21}\chi^{11/2}+{\cal
O}(\chi^{15/2})\right],\hphantom{AAAA}\label{642}
\\
\tilde{\cal Q}^4_2&=&\left( \frac{\pi}{2} \right)^\frac{1}{2}\left[
-\frac{32}{21}\chi^{9/2}-\frac{24}{77}\chi^{13/2}+{\cal
O}(\chi^{17/2})\right].\hphantom{AAAA}\label{643}
\end{eqnarray}
Hence, functions  (\ref{638})--(\ref{640}) are well behaved near the origin,
\begin{equation}
\tilde{\cal Q}^l_2\cong ({\rm const.})\chi^{l+\frac{1}{2}}~~~{\rm as}~~~\chi\rightarrow 0,
\label{644}
\end{equation}
however, they at first sight diverge at $\chi=\pi$.

Realizing that $\chi=\pi$ is an
alternative origin, we take as independent solutions (notice that Legendre
equation  is invariant under $\mu\rightarrow-\mu$)
\begin{equation}
\tilde{{\cal P} }^l_2(\chi)=\tilde{\cal Q}^l_2(\pi-\chi).
\label{645}
\end{equation}
They read
\begin{equation}
\tilde{{\cal P} }^2_2=\left( \frac{\pi}{2} \right)^\frac{1}{2}\!\frac{1}{\sin^{5/2}\chi}\left[
\frac{1}{4}\sin(4\chi)-2\sin(2\chi)-3(\pi-\chi)\right],\label{646}\\
\end{equation}
\begin{eqnarray}
\tilde{{\cal P} }^3_2&=&\left( \frac{\pi}{2} \right)^\frac{1}{2}\frac{1}{\sin^{7/2}\chi}
\Big[\frac{1}{8}\sin(5\chi)-\frac{15}{8}\sin(3\chi)-\hphantom{AAAA}
\nonumber\\
&-&10\sin\chi-15(\pi-\chi)\cos\chi\Big],
\label {647}
\end{eqnarray}
\begin{eqnarray}
\tilde{{\cal P} }^4_2&=&\left( \frac{\pi}{2} \right)^\frac{1}{2}\frac{1}{\sin^{9/2}\chi}\Big[
\frac{1}{8}\sin(6\chi)-3\sin(4\chi)-
\hphantom{AAAAA}
\nonumber\\
&-&\frac{375}{8}\sin(2\chi)-15(\pi-\chi)[3
\cos(2\chi)+4]\Big].
\label{648}
\end{eqnarray}
At $\chi\rightarrow\pi$ they behave as (\ref{641})--(\ref{643}) with $\chi\rightarrow \pi$,
so
\begin{equation}
\tilde{{\cal P} }^l_2\cong ({\rm const.})(\pi-\chi)^{l+\frac{1}{2}}~~~{\rm
as}~~~\chi\rightarrow
\pi.
\label{649}
\end{equation}

The expansions  (\ref{644}) and (\ref{649}) were obtained by direct
calculations for $l=2,3,4$. Let us briefly indicate a more general argument
demonstrating their form for general $l$. In \cite{MO} one finds (see p.178)
the recursion relations (here again $x=\cos\chi$)
\begin{equation}
xQ^\mu_\nu-Q^\mu_{\nu+1}=(\nu+\mu)(1-x^2)^\frac{1}{2}
Q^{\mu-1}_\nu
\label{650}
\end{equation}
and
\begin{equation}
(1-x^2)\frac{{\rm d}Q^\mu_\nu}{{\rm d}x}=(\nu+1)xQ^\mu_\nu-(\nu-\mu+1)
Q^\mu_{\nu+1}.
\label{651}
\end{equation}
Expressing $Q^\mu_{\nu+1}$  from the last equation, substituting into
(\ref{650}), taking $\partial/\partial\nu$ and putting $\nu=\frac{3}{2}, \mu=l+\frac{1}{2}$
and $x=\cos\chi$, one finds the recursion relation
\begin{eqnarray}
(l+\frac{1}{2})\cos\chi\frac{\partial Q^{l+\frac{1}{2}}_\nu}{\partial
\nu}\Bigg|_{3/2}+\sin\chi\frac{{\rm d}}{{\rm d}\chi}\frac{\partial Q^{l+\frac{1}{2}}_\nu}{\partial
\nu}\Bigg|_{3/2}
\nonumber\\
=(l+2)(l-2)\sin\chi\frac{\partial Q^{l-\frac{1}{2}}_\nu}{\partial
\nu}\Bigg|_{3/2}.
\label{652}
\end{eqnarray}
Let us assume that for small $\chi$ one can write
\begin{equation}
\frac{\partial Q^{l+\frac{1}{2}}_\nu}{\partial
\nu}\Bigg|_{3/2}=\alpha_l\chi^p,
\label{653}
\end{equation}
and that
\begin{equation}
\frac{\partial Q^{l-\frac{1}{2}}_\nu}{\partial
\nu}\Bigg|_{3/2}=\alpha_{l-1}\chi^{l-\frac{1}{2}}
\end{equation}
is known. Then equation   (\ref{652}) implies $p=l+\frac{1}{2}$ and
\begin{equation}
\alpha_l=\frac{(l+2)(l-2)}{l+\frac{1}{2}}\alpha_{l-1}.
\end{equation}
Continuing to $l=2$ and regarding (\ref{638}) for $\tilde{\cal Q}^2_2$, we obtain
\begin{equation}
\alpha_l=-\left( \frac{\pi}{2} \right)^\frac{1}{2}
\frac{(l+2)!(l-2)!}{(2l+1)!!},\qquad l\ge 2.
\label{654}
\end{equation}
Therefore, the asymptotic expansion (\ref{644}) can be written for general
$l\ge 2$
\begin{equation}
\tilde {\cal Q}^l_2\cong \alpha_l\chi^{l+\frac{1}{2}},
\end{equation}
where $\alpha_l$ is given by (\ref{654}). It is easy to check that for $l=2,3,4$
one gets the first terms in the expressions (\ref{641})--(\ref{643}).

Next we need to calculate the Wronskians. Defining the Wronskians with respect to $\chi$
(rather than with respect to $\mu=\cos \chi$),
\begin{equation}
{\cal W}_l\{ \tilde {\cal Q}^l_2, \tilde {{\cal P} }^l_2\}=\tilde
{\cal Q}^l_2   \frac{{\rm d}{\tilde{\cal P} }^l_2}{{\rm d}\chi}-\tilde
{\cal P} ^l_2   \frac{{\rm d}{\tilde{\cal Q}}^l_2}{{\rm d}\chi},
\label {656}
\end{equation}
and substituting from (\ref{638})--(\ref{640}) and (\ref{646})--(\ref{648}),
MATHEMATICA yields:
\begin{equation}
{\cal W}_2=-\frac{12\pi^2}{\sin\chi},~{\cal W}_3=-\frac{60\pi^2}{\sin\chi},
~{\cal W}_4=-\frac{720\pi^2}{\sin\chi}.
\label{657}
\end{equation}
Of course, the simple dependence on $\chi$ could have been obtained easily since the
Wronskians satisfy a simple differential equation. MATHEMATICA is useful to determine the
coefficients.
(If the Wronskians are defined in terms of ${\rm d}/{\rm d}\cos\chi$, the last expressions are multiplied by factor $-1/\sin\chi$.)
Playing with recurrent relations given in \cite{MO} (p.171)
for both $Q^\mu_\nu(x)$
and $Q^\mu_\nu(-x)$, and taking appropriate derivatives $\partial/\partial\nu$,
we can find the general formula for ${\cal W}_l,~~l\ge2$,  in terms of $ \tilde
{\cal Q}^l_2, \tilde {{\cal P} }^l_2$ and $Q^{l-\frac{1}{2}}_{3/2}(-x)$:
\begin{eqnarray}
{\cal W}_l=&-&(l+2)(l-2)\left(   \tilde {\cal Q}^{l-1}_2 \tilde {{\cal P} }^l_2+ \tilde {\cal Q}^l_2
\tilde {{\cal P} }^{l-1}_2 \right)
\nonumber\\
&+&4\left[  \tilde {{\cal P} }^l_2 Q^{l-\frac{1}{2}}_{3/2}(x)+
\tilde {{\cal Q}}^l_2 Q^{l-\frac{1}{2}}_{3/2}(-x)\right].
\label {658}
\end{eqnarray}
Since $Q^{l+\frac{1}{2}}_{3/2}=0$ for $l\ge2$, the second term in square
brackets does not vanish only for $l=2$.

As before, we can now solve the inhomogeneous equation  (\ref{242}) with
$k=+1,~~l=2,3,\cdots$ by variation of the parameters. Remembering that
$\tilde{\cal Q}^l_2$ is well-behaved at $\chi=0$, whereas  $\tilde {{\cal P} }^l_2$ at
$\chi=\pi$, we write the solution in the form
\begin{eqnarray}
\overline\omega_l=&&2\kappa\Bigg\{  \tilde {{\cal P} }^l_2\int_0^\chi\frac{\tilde
{{\cal Q}}^l_2}{{\cal W}_l\sin^\frac{1}{2}\chi'}(\delta T^0_\varphi)_l\,{\rm d}\chi'
\nonumber\\
&+&  \tilde
{{\cal Q}}^l_2\int_\chi^\pi\frac{\tilde
{{\cal P} }^l_2}{{\cal W}_l\sin^\frac{1}{2}\chi'}(\delta T^0_\varphi)_l\,{\rm d}\chi'\Bigg\},
\label{659}
\end{eqnarray}
where for $l=2$, $3$, $4$, functions $\tilde{\cal Q}$'s and $\tilde{\cal P} $'s are given by 
(\ref{638})--(\ref{640}) and (\ref{646})--(\ref{648}), and
$({\cal W}_l\sin^\frac{1}{2}\chi')^{-1}=({\rm const.})\sin^\frac{1}{2}\chi'$ as a
consequence of (\ref{657}). For a general $l$, $\tilde{\cal Q}^l_2=(\partial
Q^{l+\frac{1}{2}} _\nu/\partial\nu)_{\nu=3/2}$, $Q^{l+\frac{1}{2}} _\nu$ being
determined by (\ref{621}), and $\tilde{\cal P} ^l_2(\chi)=\tilde{\cal Q}^l_2(\pi-\chi)$.
The final physical solution is then obtained from $\omega_l=(\sin\chi)^{-3/2}
\overline\omega_l$, with $\overline\omega_l$ given by  (\ref{659}). It is easy to make sure
that for $\Omega-\omega$ regular at the poles (i.e., due to 
$(\delta T^0_\varphi)_l\sim\sin^2\chi$ at the poles), $\omega_l$ is well-behaved at
the poles.

Considering $(\delta T^0_\varphi)_l$ concentrated around the equator at
$\chi\in (\pi/2\pm\Delta)$, for example, then near the pole $\chi=0$ we
find
$(l\ge2)$
\begin{equation}
\omega_l=2\kappa\alpha_l\chi^{l-1}\int_{\frac{\pi}{2}-\Delta}^{\frac{\pi}{2}+\Delta}\frac{\tilde
{{\cal P} }^l_2}{{\cal W}_l\sin^\frac{1}{2}\chi'}(\delta T^0_\varphi)_l\,{\rm d}\chi',
\label{660}
\end{equation}
where factor $\alpha_l$ is given by (\ref{654}).

\subsection{The fluid angular velocity  as a source}

To solve (\ref{oml4}) with $k=+1$ and $\Omega_l$ given, we use solutions
(\ref{621}) for general $\nu=n-\frac{1}{2}$. In particular, for $l=1,\cdots,4$ we
have explicit expressions (\ref{624})--(\ref{631}) for both cases of $n$
 real and purely imaginary, $n=iN$. These solutions are well-behaved at the
origin. Indeed at small $\chi$ we find
\begin{eqnarray}
{\cal Q}^1_n&=&-\left(\frac{\pi}{2}\right)^\frac{1}{2} \frac{n}{3}(n^2-1)\chi^{3/2}+
{\cal O}(\chi^{7/2}), \label{661}\\
{\cal Q}^1_{iN}&=& i\left(\frac{\pi}{2}\right)^\frac{1}{2} \frac{N}{3}(N^2+1)\chi^{3/2}+
{\cal O}(\chi^{7/2}), \label{662}\\
{\cal Q}^2_n&=&\left(\frac{\pi}{2}\right)^\frac{1}{2} \frac{n}{15}(n^4-5n^2+4)\chi^{5/2}+
{\cal O}(\chi^{9/2}), \label{663}\\
{\cal Q}^2_{iN}&=&i\left(\frac{\pi}{2}\right)^\frac{1}{2}
\frac{N}{15}(N^4+5N^2+4)\chi^{5/2}\!+\!{\cal O}(\chi^{9/2}), \label{664}\\
{\cal Q}^3_n&=&-\hphantom{i}\left(\frac{\pi}{2}\right)^\frac{1}{2}
\frac{n}{105}(-n^6+14n^4-49n^2+36)\chi^{7/2}+
\nonumber\\&&+ {\cal O}(\chi^{11/2}),
\label{665}\\
{\cal Q}^3_{iN}&=&-i\left(\frac{\pi}{2}\right)^\frac{1}{2}
\frac{N}{105}(N^6+14N^4+49N^2+36)\chi^{7/2}+
\nonumber\\&&+ {\cal O}(\chi^{11/2}),
\label{666}\\
{\cal Q}^4_n&=&-\hphantom{i}\left(\frac{\pi}{2}\right)^\frac{1}{2}
\frac{n}{945}(n^8-30n^6+273n^4-820n^2+
\nonumber\\&&+576)\chi^{9/2}+ {\cal
O}(\chi^{13/2}),
\label{667}\\
{\cal Q}^4_{iN}&=&-i\left(\frac{\pi}{2}\right)^\frac{1}{2}
\frac{N}{945}(N^8+30N^6+273N^4+820N^2+
\nonumber\\&&+576)\chi^{9/2}+ {\cal
O}(\chi^{13/2}).
\label{668}
\end{eqnarray}
To obtain these expansions we could not have used formulas given, for
example, in \cite{MO} (see p.196), since the leading term in
$Q^\mu_\nu(x)$ at $x=1$ is proportional to $\cos(\pi\mu)$; hence, it
vanishes for $\mu=l+\frac{1}{2}$. However, we may proceed similarly to the
way how we obtained the coefficients $\alpha_l$ in (\ref{654}). We substitute
for $Q^\mu_{\nu+1}(x)$ in (\ref{650}) from equation  (\ref{651}), put
$x=\cos\chi, ~ \mu=l+\frac{1}{2},~ \nu=n-\frac{1}{2}, ~
Q^{l+\frac{1}{2}} _{n-\frac{1}{2}}={\cal Q}^l_n$, and so arrive at
the recursion relation
\begin{equation}
{\cal Q}^l_n\frac{l+\frac{1}{2}}{n-l}+\tan\chi\frac{1}{n-l}\frac{{\rm d}{\cal Q}^l_n}{{\rm d}\chi}+
(l+n)\tan\chi{\cal Q}^{l-1}_n=0.
\label{669}
\end{equation}
Regarding results  (\ref{661})--(\ref{668}), assume that for small $\chi$
\begin{equation}
{\cal Q}^l_n=\alpha_{l,n}\chi^p,
\label {670}
\end{equation}
and that ${\cal Q}_n^{l-1}=\alpha_{l-1,n}\chi^{l-\frac{1}{2}}$
is known. Then (\ref{669}) implies $p=l+\frac{1}{2}$ and
\begin{equation}
\alpha_{l,n}=- \frac{(n+l)(n-l)}{2l+1}\alpha_{l-1,n}.
\end{equation}
Continuing to $l=1$ we get
\begin{eqnarray}
\alpha_{l,n}=\phantom{AAAAA}&\!\!\!&\!\!\!
\\\nonumber
(-1)^l(\frac{\pi}{2})^\frac{1}{2}
\frac{n(n^2-l^2)[n^2-(l-1)^2]\cdots(n^2-4)(n^2-1)}{(2l+1)!!}.&\!\!\!&\!\!\!
\label{671}
\end{eqnarray}
For $n=iN$,
\begin{equation}
{\cal Q}^l_{iN}=\alpha_{l,iN}\chi^{l+\frac{1}{2}},
\label{672}
\end{equation}
where
\begin{eqnarray}
&&\alpha_{l,iN}=i\left(\frac{\pi}{2}\right)^\frac{1}{2}
\times\\\nonumber
&\times&\frac{N(N^2+l^2)[N^2+(l-1)^2]\cdots(N^2+4)(N^2+1)}{(2l+1)!!}.
\label{673}
\end{eqnarray}

 The second (independent) solutions which are well-behaved at the other
pole, $\chi=\pi$, are, as before, obtained by taking ${\cal Q}$'s at $\pi-\chi$.
Thus,
\begin{equation}
{\cal P} ^l_n(\chi)={\cal Q}^l_n(\pi-\chi)~~~,~~~{\cal P} ^l_{iN}(\chi)=
{\cal Q}^l_{iN}(\pi-\chi)
\label{674}
\end{equation}
are solutions to the homogeneous Legendre equation  (\ref{620}) on the cut
which are regular at $\chi=\pi$.

We do not write them down explicitly since they follow easily from ${\cal Q}$'s
given in (\ref{624})--(\ref{631}). At $\chi\rightarrow\pi$
\begin{equation}
{\cal P} ^l_n=\alpha_{l,n}(\pi-\chi)^{l+\frac{1}{2}}~~~~,~~~~
{\cal P} ^l_{iN}=\alpha_{l,iN}(\pi-\chi)^{l+\frac{1}{2}}\;.
\label{675}
\end{equation}

What are the Wronskians? We get
\begin{eqnarray}
{\cal W}_{1,n}\{{\cal Q}^1_n,{\cal P} ^1_n\}&\!=\!&\hphantom{-}\frac{\pi}{2}n(n^2-1)
\frac{\sin(n\pi)}{\sin\chi},\label{676}\\
{\cal W}_{1,N}\{{\cal Q}^1_{iN},{\cal P} ^1_{iN}\}&\!=\!&\hphantom{-}\frac{\pi}{2}N(N^2+1)
\frac{\sinh(N\pi)}{\sin\chi},\label{677}\\
{\cal W}_{2,n}\{{\cal Q}^2_n,{\cal P} ^2_n\}&\!=\!&-\frac{\pi}{2}n(n^4\!-5n^2\!+4)
\frac{\sin(n\pi)}{\sin\chi},\label{678}\\
{\cal W}_{2,N}\{{\cal Q}^2_{iN},{\cal P} ^2_{iN}\}&\!=\!&\hphantom{-}\frac{\pi}{2}N(N^4+5N^2+4)
\times\hphantom{AAAAAA}
\nonumber\\
&&~~~~\times\frac{\sinh(N\pi)}{\sin\chi},\label{679}
\end{eqnarray}
where we restricted ourselves to $l=1,2$ since 
the polynomials depending on $n$ (resp.\,$N$) are proportional to those in the
expansions  ~~~(\ref{661})--(\ref{668}) at $\chi\rightarrow0$ for which we
succeeded to derive the formulas (\ref{671})--(\ref{673}) for general $l$. 
(This  is due to the fact that the functional dependence of the Wronskian is governed by a
simple differential equation, and the proportionality factor can be determined at any
point, e.g., at $\chi\rightarrow 0$.)
The general form of the Wronskians can be derived by expressing functions 
${\cal P} ^l_n$ in terms of $Q^\mu_\nu$'s and $P^\mu_\nu$'s. Using the
relation giving $Q^\mu_\nu(-x)$ in terms of $Q^\mu_\nu(x)$ and
$P^\mu_\nu(x)$ [not to be confused with
${\cal P} ^l_n={\cal Q}^\mu_\nu(\pi-\chi)$] -- see e.g.
\cite{MO}, p.170 -- and putting $\mu=l+\frac{1}{2},~\nu=n-\frac{1}{2}$, we get
\begin{eqnarray}
{\cal P} _n^l&=&Q^{l+\frac{1}{2}}_{n-\frac{1}{2}}(-x)=(-1)^{l+1}
\times
\\\nonumber
&\times&\left[
Q^{l+\frac{1}{2}}_{n-\frac{1}{2}}(x)\cos(n\pi)+\frac{\pi}{2}
P^{l+\frac{1}{2}}_{n-\frac{1}{2}}(x)\sin(n\pi)\right].
\label{684}
\end{eqnarray}
Employing then the well known formula (\cite{MO}, p.170)
\begin{equation}
{\cal W}\left[   P^\mu_\nu, Q^\mu_\nu\right]=
\frac{\Gamma(1+\nu+\mu)}{\Gamma(1+\nu-\mu)}\frac{1}{1-x^2},
\label{685}
\end{equation}
we arrive at
\begin{eqnarray}
{\cal W}_{l,n}[{\cal Q}^l_n,{\cal P} ^l_n]&=&(-1)^{l+1}\frac{\pi}{2}n(n^2-l^2)
\times
\label{686}\\\nonumber
\times[n^2-(l-1)^2]&\cdots&(n^2-4)(n^2-1)~\frac{\sin(n\pi)}{\sin\chi},
\end{eqnarray}
and
\begin{eqnarray}
{\cal W}_{l,N}[{\cal Q}^l_{iN},{\cal P} ^l_{iN}]&=&\frac{\pi}{2}N(N^2+l^2)\times
\label{687}\\\nonumber\times[N^2
+(l-1)^2]&\cdots&(N^2+4)(N^2+1)\times\frac{\sinh(N\pi)}{\sin\chi}.
\end{eqnarray}
It is easy to regain  (\ref{676})--(\ref{679}) for $l=1,2$.

Now we can finally solve the inhomogeneous equation  (\ref{oml4}), with
$\Omega_l$ as a source, again by the variation of the parameters. Regarding
the relation  (\ref{241}) between $\overline\omega_l$ and $\omega_l$ we arrive at
the solutions in the form
\begin{eqnarray}
\omega_l=\frac{n^2-4}{(\sin\chi)^{3/2}}\Bigg\{ {\cal P} ^l_n\int_0^\chi
\frac{(\sin\chi')^{3/2}}{{\cal W}_{l,n}}{\cal Q}^l_n\Omega_l\,{\rm d}\chi'+
\nonumber\\
 {\cal Q}^l_n\int_\chi^\pi
\frac{(\sin\chi')^{3/2}}{{\cal W}_{l,n}}{\cal P} ^l_n\Omega_l\,{\rm d}\chi'\Bigg\},\hphantom{AAAA}
\label{688}
\end{eqnarray}
analogously for $n=iN$:
\begin{eqnarray}
\omega_l=-\frac{N^2+4}{(\sin\chi)^{3/2}}\Bigg\{ {\cal P} ^l_{iN}\int_0^\chi
\frac{(\sin\chi')^{3/2}}{{\cal W}_{l,N}}{\cal Q}^l_{iN}\Omega_l\,{\rm d}\chi'+
\nonumber\\
+ {\cal Q}^l_{iN}\int_\chi^\pi
\frac{(\sin\chi')^{3/2}}{{\cal W}_{l,N}}{\cal P} ^l_{iN}\Omega_l\,{\rm d}\chi'\Bigg\}.\hphantom{AAAA}
\label{689}
\end{eqnarray}
All quantities entering (\ref{688}) and (\ref{689})
are given above (for $l=1,\cdots, 4$  in the explicit forms), the
time-dependent `constants' $n, N$ are determined by (\ref{244}), which
was used to express $\lambda^2$ in terms of $n^2$ and $N^2$. Notice that
solutions (\ref{689})
are all real since ${\cal Q}^l_{iN}$ and ${\cal P} ^l_{iN}$ are purely imaginary.
We could have divided
all solutions of the homogeneous equation  given in the present section by $n$
(resp.$N$) as we did for $l=1$ in Paper~I, and so define
$^\ast{\cal Q}^l_n={\cal Q}^1_n/n,~
^\ast{\cal Q}^l_{iN}={\cal Q}^1_{iN}/N$ etc which, as can easily be seen, remain
non-vanishing for $n=0, N=0$.  However, the Wronskians (\ref{686}) and
(\ref{687}) are proportional to $n^2$ (resp.$N^2$) as $n\rightarrow 0$ 
(resp.$N\rightarrow 0$), so the expressions ${\cal Q}_n^l{\cal P} _n^l/{\cal W}_{l,n}$ (the same with
$N$) remain non-vanishing in this limit. Hence physical solutions
(\ref{688}), (\ref{689}) for $\omega_l$ do not vanish when $n=0$
(resp. $N=0$).

Now the final solutions (\ref{688}), (\ref{689}) for any $l$
can be applied to the special situations considered for $l=1$ in Paper~I.
So, for large $N$ (\ref{687}) implies
\begin{equation}
{\cal W}_{l,N}=\frac{\pi}{2}N^{2l+1}\frac{e^{N\pi}}{2\sin\chi},
\label{690}
\end{equation}
and the inspection of (\ref{624})--(\ref{631}) and the relation (\ref{675})
lead to
\begin{equation}
{\cal P} ^l_{iN}=i(-1)^l\left( \frac{\pi}{2} \right)^\frac{1}{2}\frac{1}{2}\frac{N^le^{N(\pi-\chi)}}
{\sin^{1/2}\chi}.
\label{691}
\end{equation}
For $N$ large but $\chi$ small expansions (\ref{661})--(\ref{668}) and
general formulas (\ref{671}), (\ref{673}) give
\begin{equation}
{\cal Q}^l_{iN}=i\left( \frac{\pi}{2} \right)^\frac{1}{2}\frac{N^{2l+1}}{(2l+1)!!}\chi^{l+\frac{1}{2}}.
\label{692}
\end{equation}
Hence our solution  (\ref{689}) for $N$ large near the origin is
\begin{equation}
\omega_l=\frac{N^{l+2}\chi^{l-1}}{(2l+1)!!}\int_\chi^\pi\Omega
_l(\chi')\sin^2\chi'e^{-N\chi'}\,{\rm d}\chi'.
\label{693}
\end{equation}
For $l=1$ we recover the leading term (independent of $\chi$) in the
expression (I.4.28).  The next term in front of the integral in (\ref{693}),
which is $\sim \chi^{l+1}$, comes from the second terms in the expansions
(\ref{661})--(\ref{668})
indicated only by ${\cal O}(\chi^{l+\frac{5}{2}})$; their exact forms, if
needed, can be obtained without difficulty by MATHEMATICA. The resulting
$\omega_l$ at given small $\chi$ can become large at high $l$'s since
$N^2=\kappa a^2(\rho+p)-4$ may become very large in, say, radiation
dominated universes (when $(\rho+p)\sim a^{-4}$) close to big bang. Near
the perturbation  with $\Omega_l$ concentrated in $r_0(\chi_0)\pm\Delta$, with
$N\Delta\ll1
, N$ large, the solution   (\ref{689})
leads to
\begin{equation}
\omega_l(\chi_0)=\frac{1}{2}\int_0^\pi Ne^{-N|\chi'-\chi_0|}\Omega_l(\chi')\left(
\frac{\sin\chi'}{\sin\chi_0} \right)^2\,{\rm d}\chi',
\label{694}
\end{equation}
i.e., the result identical for all $l$, and thus coinciding for $l=1$ with
(I.4.29), in which $\lambda^2=N^2+4$, and only the leading term in $N$
is kept.


\vskip .5 cm
\sect{Axisymmetric toroidal perturbations for $k=-1$}
\vskip .5 cm

With $k=-1$, the homogeneous differential equation  corresponding to the equation 
(\ref{oml4}) with $\Omega_l$ as a source takes the form
\begin{eqnarray}
\frac{\partial}{\partial\mu}\left[
(1-\mu^2)\frac{\partial\overline\omega_l}{\partial\mu}\right]+\left[ \nu(\nu+1)-
\frac{(l+\frac{1}{2})^2}{1-\mu^2}\right]\overline\omega_l=0,
\label{695}&~~~&
\end{eqnarray}
again the Legendre equation. Now $\mu=\sqrt{1+r^2}=\cosh\chi,\,
r=\sinh\chi$, so the independent variable $\mu\in\left<1,+\infty\right)$. The
solutions of the Legendre equation  for $\mu$ such that its real part is $\ge1$
($\mu$ is then often written as $z$) are sometimes denoted by capital
Gothic script letters. This is the case in \cite{MO} from which we shall again
take some formulas but we keep the Latin script. If $\nu=\frac{3}{2}$, equation 
(\ref{695})
 corresponds to the inhomogeneous equation  (\ref{242})
with $(\delta T^0_\varphi)_l$ as a source.

As in the case $k=+1$, we start from the general expression for the solution
in terms of the combination of two hypergeometric functions. We first
consider the Legendre function  $P^\mu_\nu(z)$
as given in \cite{MO} (p. 154, the fourth line), substitute for
$z=\mu=\cosh\chi$, put $\mu=l+\frac{1}{2}$ and, as before, $\nu+\frac{1}{2}=n$.
Regarding standard properties of the hypergeometric functions for special
values of parameters, we find after simple algebra that functions 
$P^{l+\frac{1}{2}}_{n-\frac{1}{2}}(\cosh\chi)$ can be rewritten as the following finite sums:
\begin{widetext}
\begin{eqnarray}
P^{l+\frac{1}{2}}_{n-\frac{1}{2}}&=&\frac{1}{\sqrt{2\pi}}
\frac{\Gamma(n+l+1)}{\Gamma(n+1)}\frac{1}{(\sinh\chi)^\frac{1}{2}}
\Bigg\{\sum\limits_{k=0}
^l \frac{(-l)_k(l+1)_k}{(n+1)_kk!}\left[  \frac{1}{(2\sinh\chi)^k}\left(
e^{(k+n)\chi}+ (-1)^{l+k}e^{-(k+n)\chi}    \right)\right]\Bigg\},
\label{696}
\end{eqnarray}
(cf. (\ref{621})). We shall also use
\begin{eqnarray}
P^{-(l+\frac{1}{2})}_{n-\frac{1}{2}}&=&\frac{1}{\sqrt{2\pi}}
\frac{\Gamma(n-l)}{\Gamma(n+1)}\frac{1}{(\sinh\chi)^\frac{1}{2}}
\Bigg\{\sum\limits_{k=0}
^l \frac{(-l)_k(l+1)_k}{(n+1)_kk!}\left[  \frac{1}{(2\sinh\chi)^k}\left(
e^{(k+n)\chi}- (-1)^{l+k}e^{-(k+n)\chi}\right)\right]\Bigg\},
\label{697}
\end{eqnarray}
\end{widetext}
which is also the solution since the Legendre equation  is invariant under the
change $l+\frac{1}{2}\rightarrow -(l+\frac{1}{2})$. This solution can be written as a linear
combination of $P^{l+\frac{1}{2}}_{n-\frac{1}{2}}$ and of another solution,
$Q^{l+\frac{1}{2}}_{n-\frac{1}{2}}$, as follows\footnote{For a formula connecting these
solutions for general parameters in the Legendre equation, see \cite{MO}, p.
164.}
\begin{equation}
P^{-(l+\frac{1}{2})}_{n-\frac{1}{2}}=\frac{\Gamma(n-l)}{\Gamma(n+l+1)}\left(   P^{l+\frac{1}{2}}_{n-\frac{1}{2}}
+\frac{2}{\pi}iQ^{l+\frac{1}{2}}_{n-\frac{1}{2}}\right).
\label{698}
\end{equation}
The last equation  combined with (\ref{696}) and (\ref{697}) can be
used to find a suitable form of the solutions $Q^{l+\frac{1}{2}}_{n-\frac{1}{2}}$:
\begin{eqnarray}
Q^{l+\frac{1}{2}}_{n-\frac{1}{2}}&=&i\left(\frac{\pi}{2}
\right)^\frac{1}{2}\frac{(n+l)(n+l-1)\cdots(n+1)}{(\sinh\chi)^{\frac{1}{2}}}
\times
\nonumber\\
&\times&
\sum\limits_{k=0}
^l \frac{(-1)^{l+k}(-l)_k(l+1)_k}{(n+1)_kk!}\frac{e^{-(k+n)\chi}}
{(2\sinh\chi)^k}.
\label{699}
\end{eqnarray}
Another useful expression for $P^{-\nu-\frac{1}{2}}_{-\mu-\frac{1}{2}}$ is provided by
the inversion of Whipple's formula (see e.g. \cite{MO}, p. 164) which in our
case of $\mu$ and $\nu$ reads
\begin{equation}
P^{-(l+\frac{1}{2})}_{n-\frac{1}{2}}\left(  \frac{z}{\sqrt{z^2-1}}\right)=
\frac{(z^2-1)^{\frac{1}{4}}e^{in\pi}Q^{-n}_l (z)}{\sqrt{\pi/2}~\Gamma(l-n+1)}\;,
\label{6100}
\end{equation}
where  $z=\cosh\chi$ (notice that the orders and degrees in $P$ and $Q$ are
here interchanged).

The case $k=-1$ is simpler than that with $k=+1$. 
The ``constant'' parameter $n=\nu+1/2$ does not become imaginary (cf. equation  (\ref{244})),
and the particular case $\nu=3/2, ~n=2$, corresponding to the
homogeneous equation  in (\ref{242}) with $\delta T^0_\varphi$ as a source, does
not require as special procedure as for $k=+1$. The relevant independent
solutions of the homogeneous equations  are
\begin{equation}
\overline{\cal P} ^l_n=P^{-(l+\frac{1}{2})}_{n-\frac{1}{2}}~~~,~~~\overline{\cal Q}^l_n=-iQ^{l+\frac{1}{2}}_{n-\frac{1}{2}},
\label{6101}
\end{equation}
where we use the notation similar to that with $k=+1$ but put bars above
${\cal P} $'s and ${\cal Q}$'s to distinguish the two cases. The solutions
(\ref{6101}) are given as finite sums by  (\ref{697}),  (\ref{699}),
respectively  (\ref{6100}).

The asymptotic behavior of $\overline{\cal Q}^l_n$ and $\overline{\cal P} ^l_n$ at the
origin and at infinity can be read off from literature where asymptotic forms
of $P^\mu_\nu$ and $Q^\mu_\nu$  are given (e.g.
\cite{MO}, p. 196 and 197). After arrangements we find at small $\chi$
\begin{eqnarray}
\overline{\cal P} ^l_n&=&\left(  \frac{2}{\pi} \right)^\frac{1}{2}
\frac{\chi^{l+\frac{1}{2}}}{(2l+1)!!},\label{6102}\\
\overline{\cal Q}^l_n&=& (-1)^{l+1} \left(  \frac{\pi}{2} \right)^\frac{1}{2} 
\frac{(2l-1)!!}{\chi^{l+\frac{1}{2}}},
\label{6103}
\end{eqnarray}
at large $\chi$
\begin{eqnarray}
\overline{\cal P} ^l_n&=&\frac{1}{\sqrt{\pi}}
\frac{(2\cosh\chi)^{n-\frac{1}{2}}}{(n+l)(n+l-1)\cdots n},\label{6104}\\
\overline{\cal Q}^l_n&=&(-1)^l \sqrt{\pi}
\frac{(n+l)(n+l-1)\cdots(n+1)}{(2\cosh\chi)^{n+\frac{1}{2}}}.\hphantom{AAAA}
\label{6105}
\end{eqnarray}

Before we consider the solutions of the inhomogeneous equations  we list explicit
expressions for $\overline{\cal P} ^l_n$  and $\overline{\cal Q}^l_n$ for $l=1,2,3$ as they
follow, after suitable arrangements, from general formulas (\ref{697}),
(\ref{699}) and the definition (\ref{6101}). They look as follows:
\begin{widetext}
\begin{eqnarray}
\overline{\cal P} ^1_n&=&\left(  \frac{2}{\pi}
\right)^\frac{1}{2}\frac{1}{2n(n^2-1)(\sinh\chi)^{3/2}}\Big\{
(n-1)\sinh[(n+1)\chi]-(n+1)\sinh[(n-1)\chi]
\Big\},\nonumber\\  \label{6106}\\
\overline{\cal Q}^1_n&=&-\left(  \frac{\pi}{2} \right)^\frac{1}{2}\frac{1}{2(\sinh\chi)^{3/2}}
\left[   (n+1) e^{-(n-1)\chi} - (n-1) e^{-(n+1)\chi}\right],\label{6107}\\
\overline{\cal P} ^2_n&=&\left(  \frac{2}{\pi}
\right)^\frac{1}{2}\frac{1}{n(n^2-1)(n^2-4)(\sinh\chi)^{5/2}}\bigg\{
(n+2)\sinh\chi[-3\cosh\{ (n+1)\chi  \}
\nonumber\\&&~~~~~~~~~~~~~~~~~~~~~~~~~~~~~~~~~~~~
+(n+1)\sinh\chi\sinh(n\chi)]+3\sinh[(n+2)\chi]\bigg\},
   \label{6108}\\
\overline{\cal Q}^2_n&=&\left(  \frac{\pi}{2} \right)^\frac{1}{2}\frac{1}{2(\sinh\chi)^{5 /2}}
\Bigg\{   (n+2) \sinh\chi\left[ (n+1)\sinh\chi
e^{-n\chi}+3e^{-(n+1)\chi}\right]
+3e^{-(n+2)\chi}\Bigg\}\;, \nonumber \\
\label{6109}\\
\overline{\cal P} ^3_n&=&\left(  \frac{2}{\pi}
\right)^\frac{1}{2}\frac{1}{n(n^2-1)(n^2-4)(n^2-9)(\sinh\chi)^{7/2}}\bigg\{
(n+3)\sinh\chi\Bigl[15\cosh\{ (n+2)\chi  \}\nonumber\\
&+&(n+2)\sinh\chi\left[
(n+1)\cosh(n\chi)\sinh(n\chi)-6\sinh\{(n+1)\chi\}\right]\Bigr]\!-\kern-.6pt 15\sinh\{
(n+3)\chi
\}\bigg\},\nonumber\\
\label{6110}\\
\overline{\cal Q}^3_n&=&-\left(  \frac{\pi}{2} \right)^\frac{1}{2}\frac{1}{(\sinh\chi)^{7 /2}}
\Bigg\{   (n+3) \sinh\chi\bigg [
15e^{-(n+2)\chi}+(n+2)\sinh\chi \times\nonumber\\
&&~~~~~~~~~~~~~~~~~~~~~~~~~~~~~~\times{\left[
(n+1)\sinh\chi e^{-n\chi}+6e^{-(n+1)\chi}\right]} \bigg]+15
e^{-(n+3)\chi}\Bigg\}.\label{6111}
\end{eqnarray}
\newpage
\end{widetext}

\subsection{The angular momentum  as a source of $\omega$}

We need to solve the homogeneous equation  (\ref{695})
for $\nu=\frac{3}{2},~n=2$.  At first sight we again encounter a peculiar case
as with $k=+1$, since $\Gamma(n-l)$ in (\ref{697}) has a pole at $l=2$ for
$n=2$, the expression (\ref{698}) appears to have also a pole in this case,
and explicit formulas (\ref{6108}) and (\ref{6110}) seem to diverge for
$n=2$. 
However, Whipple's formula (\ref{6100}) for $\overline{\cal P} ^l_n$ and
the following formula
\begin{equation}
\overline{\cal P} ^l_n(\tilde z) = \frac{1}{\sqrt{\pi/2}}
\frac{e^{-in\pi}}{\Gamma(l+n+1)}\frac{1}{({\tilde z}^2-1)^{1/4}}Q^n_l(\tilde z/\sqrt{\tilde
z^2-1}),
\label{6112}
\end{equation}
which, after employing the relation between $Q^{-n}_l$ and $Q^n_l$ \cite{MO},
is its consequence, show no peculiar behavior for $n=2$. In fact,
the last formula can be used to obtain simple explicit expressions for
$\overline{\cal P} ^l_n$ because (e.g. \cite{AS}, formula 8.6.7)
\begin{equation}
Q^2_l(z)= (z^2-1)\frac{d^2Q_l}{dz^2},
\label{6113}
\end{equation}
where
\begin{equation}
Q_l(z)= \frac{1}{2} P_l(z)\log\frac{z+1}{z-1}+W_{l-1}(z),
\label{6114}
\end{equation}
$W_{l-1}$ are polynomials of degree $l-1$ given in terms of Legendre
polynomials $P_l(z)$.

The same result can be derived directly from the original expressions
(\ref{697}) and (\ref{699})
by taking the limit $n\rightarrow2$. So, we take as our solutions of the homogeneous  equations 
(\ref{695}) for $\nu=3/2$, $n=2$ the functions 
\begin{equation}
\overline{\cal P} _2^l=\lim\limits_{n\rightarrow2}\overline{\cal P} _n^l,
\qquad \overline{\cal Q}_2^l=\lim\limits_{n\rightarrow2}\overline{\cal Q}_n^l,
\label{6115}
\end{equation}
where $\overline{\cal P} _n^l$ and $\overline{\cal Q}_n^l$
are given by (\ref{6101}), (\ref{697}) and (\ref{699}). The behavior of
$\overline{\cal P} _2^l$ and $\overline{\cal Q}_2^l$ at $\chi\rightarrow\infty$
is given by  (\ref{6104}) and (\ref{6105}) with $n=2$; the first term in the 
expansions of $\overline{\cal P} _n^l$ and $\overline{\cal Q}_n^l$ at $\chi\rightarrow0$, equations 
(\ref{6102}) and  (\ref{6103}), does not depend on   $n$, so $\overline{\cal P} _2^l$
at $\chi\rightarrow0$ behaves as $\overline{\cal P} _n^l$.

The explicit expressions for $l=1,2,3$ follow easily from taking the limits
$n\rightarrow2$ in  (\ref{6106})--(\ref{6111}). One finds
\begin{eqnarray}
\overline{\cal P} ^1_2&=&\hphantom{-}\left(  \frac{2}{\pi} \right)^\frac{1}{2}
\frac{1}{3}(\sinh\chi)^{3/2},\label{6116}\\
\overline{\cal Q}^1_2&=&- \left(\frac{\pi}{2} \right)^\frac{1}{2}\frac{1}{(\sinh\chi)^{3/2}}\big[
\cosh\chi(1-2\sinh^2\chi)+
\nonumber\\&~&
+2\sinh^3\chi \big],\label{6117}\\
\overline{\cal P} ^2_2&=&\hphantom{-}\left(  \frac{2}{\pi} \right)^\frac{1}{2}
\frac{1}{(\sinh\chi)^{5/2}}\frac{1}{96}\big[  12\chi-8\sinh(2\chi)+
\nonumber\\&~&+\sinh(4\chi)
\big],\label{6118}
\\
\overline{\cal Q}^2_2&=& \hphantom{-}\left(\frac{\pi}{2} \right)^\frac{1}{2}\frac{3}{(\sinh\chi)^{5/2}},
\label{6119}
\end{eqnarray}
\begin{eqnarray}
\overline{\cal P} ^3_2&=&\hphantom{-}\left(  \frac{2}{\pi} \right)^\frac{1}{2}
\frac{1}{(\sinh\chi)^{7/2}}\frac{1}{480}\big[
120\chi\cosh\chi-
\nonumber\\&~&
-80\sinh\chi-15\sinh(3\chi)+\sinh(5\chi)
\big],
\label{6120}\\
\overline{\cal Q}^3_2&=&- \left(\frac{\pi}{2} \right)^\frac{1}{2}\frac{1}{(\sinh\chi)^{7/2}}\Big\{
5\sinh\chi\big[  e^{-4\chi}+
\\\nonumber&~&
+4\sinh\chi\left(   3\sinh\chi
e^{-2\chi}+6e^{-3\chi}\right) \big]  +15e^{-5\chi}
\Big\} .
\label{6121}
\end{eqnarray}

Starting from the well-known formula (e.g. \cite{MO}, p.165) for the
Wronskian ${\cal W}\{ P^{-\mu}_\nu,Q^\mu_\nu \}=e^{i\pi\mu}/(1-z^2)$,
$z=\cosh\chi$, which is independent of $\nu$ and, in our
case with $\mu=l+\frac{1}{2}$, the value of $l$ changes only the sign of ${\cal W}$, we
easily get
\begin{equation}
\overline{\cal W}_{l,n}\{ \overline{\cal P} ^l_n,\overline{\cal Q}^l_n  \}=\overline{\cal W}_{l}\{ \overline{\cal P} ^l_2
,\overline{\cal Q}^l_2  \}=\frac{(-1)^{l+1}}{\sinh\chi}.
\label{6122}
\end{equation}

Employing again the method of variation of parameters and regarding the
asymptotic behavior of $ \overline{\cal P} ^l_n $ and $\overline{\cal Q}^l_n  $, we arrive at
the well-behaved solutions of the inhomogeneous  equation  (\ref{oml4}) -- and hence of
the original equation   (\ref{241}) -- in the form
\begin{eqnarray}
\omega_l&=&\frac{2\kappa}{(\sinh\chi)^{3/2}}\Bigg[
\overline{\cal Q}^l_2\int_0^\chi\frac{\overline{\cal P} ^l_2(\delta
T^0_\varphi)_l}{\overline{\cal W}_l(\sinh\chi')^\frac{1}{2}}\,{\rm d}\chi'+
\nonumber\\&~&
+  \overline{\cal P} ^l_2\int_\chi^
\infty\frac{\overline{\cal Q}^l_2(\delta
T^0_\varphi)_l}{\overline{\cal W}_l(\sinh\chi')^\frac{1}{2}}\,{\rm d}\chi'
\Bigg].
\label{6123}
\end{eqnarray}

For $l=1$ the angular momentum  in $\left<0,\chi\right>$ is given by equation  (\ref{634}), with
$\sin\chi'\rightarrow\sinh\chi'$; $\omega_1$ and $(\delta T^0_\varphi)_1$ 
are connected with $\omega$ and $\delta T^0_\varphi$ as in the cases
$k=0,+1$. Noticing that $\overline{\cal W}$
 defined in (I.3.13), is related to $\overline{\cal Q}^1_2$ given in  (\ref{6117})
by $\overline{\cal W}=-\sqrt{2/\pi}(\sinh\chi)^{-3/2}\overline{\cal Q}^1_2$, we easily find
that for $l=1$ the solution (\ref{6123})
 goes over into the solution (I.3.14).

Next consider $(\delta T^0_\varphi)_l$ concentrated at $\chi\in(\chi_0\pm\Delta)$,
 $\chi_0>0$. Then near the origin $\chi\rightarrow0$
\begin{equation}
\omega_l=2\kappa\overline\alpha_l\chi^{l-1}\int_{\chi_0-\Delta}^{\chi_0+\Delta}
\overline{\cal Q}^l_2(-1)^{l+1} (\sinh\chi')^\frac{1}{2}(\delta T^0_\varphi)_l \,{\rm d}\chi',
\label{6124}
\end{equation}
where (see (\ref{6102}))
\begin{equation}
\overline\alpha_l=\left(   \frac{2}{\pi}\right)^\frac{1}{2}\frac{1}{(2l+1)!!}.
\label{6125}
\end{equation}

\subsection{The fluid angular velocity  as a source}

To solve the inhomogeneous  equation   (\ref{oml4}) with $k=-1$ and $\Omega_l$ given, we
use solutions $ \overline{\cal P} ^l_n $ and $\overline{\cal Q}^l_n  $  determined by
(\ref{6101}),  (\ref{697}) and  (\ref{699}). For $l=1,2,3$ they are given in
explicit, arranged forms by  (\ref{6106})--(\ref{6111}). The solution
$\overline\omega_l$, obtained by variation of the parameters, leads, regarding
(\ref{242}), to the following physical solution well-behaved at the origin
and vanishing at infinity:
\begin{eqnarray}
\omega_l&=&\frac{1}{(\sinh\chi)^{3/2}}\Bigg[
\overline{\cal Q}^l_n\int_0^\chi\frac{\overline{\cal P} ^l_n(-\lambda^2\Omega_l)(\sinh\chi')^{3/2}}
{\overline{\cal W}_{l,n}}\,{\rm d}\chi'+
\nonumber\\&~&
+ \overline{\cal P} ^l_n\int_\chi^\infty\frac{\overline
{\cal Q}^l_n(-\lambda^2\Omega_l)(\sinh\chi')^{3/2}}
{\overline{\cal W}_{l,n}}\,{\rm d}\chi'
\Bigg].
\label{6126}
\end{eqnarray}
Substituting for $\overline{\cal W}_{l,n}$
from (\ref{6122}) and for $\lambda^2(t)=n^2(t)-4$ as it follows from
(\ref{244}) with $k=-1$, we finally have
\begin{eqnarray}
\omega_l&=&\frac{(-1)^l(n^2-4)}{(\sinh\chi)^{3/2}}\Bigg\{   \overline{\cal Q}^l_n\int_0^\chi
 \overline{\cal P} ^l_n\Omega_l(\sinh\chi')^{5/2}\,{\rm d}\chi'+
\nonumber\\&~& 
+   \overline{\cal P} ^l_n\int_\chi^\infty
 \overline{\cal Q}^l_n\Omega_l(\sinh\chi')^{5/2}\,{\rm d}\chi'     \Bigg\}.
\label{6127}
\end{eqnarray}
For $l=1$ we notice that with 
$\overline{\cal Q}^1_n$ and $\overline{\cal P} ^1_n$ we regain the
solution (I.4.34).

Assume $\Omega_l$ non-vanishing at large $\chi$. Then the solution
(\ref{6127}) and the asymptotic results for $\overline{\cal Q}^1_n$ and
$\overline{\cal P} ^1_n$, equations  (\ref{6102})--(\ref{6105}), imply that at small $\chi$
\begin{eqnarray}
\omega_l&=&\frac{   (n^2-4)(n+l)(n+l-1)\cdots(n+1)  }{ 4(2l+1)!!  }\chi^{l-1}
\times
\nonumber\\&~&
\times
\int_\chi^\infty
(\sinh\chi')^2e^{-n\chi'}\Omega_l\,{\rm d}\chi'.
\label{6128}
\end{eqnarray}
Close to the perturbation,
\begin{equation}
\omega_l=\frac{1}{2}\frac{   (n^2-4)  }{n }
\int_0^\infty
\left(\frac{\sinh\chi'}{\sinh\chi_0}\right)^2e^{-n|\chi'-\chi_0|}\Omega_l\,{\rm d}\chi'.
\label{6129}
\end{equation}
Here $l$ does enter the result directly. So for $l=1$ it goes over immediately
to (I.4.38). It is also easy to see that (\ref{6128})  gives for $l=1$ the
first terms in (I.4.37).

\begin{acknowledgements}
This work started during our meeting at the Institute of Theoretical Physics
of the Charles University in Prague and continued during our stay at the
Albert Einstein Institute in Golm. We are grateful to these Institutes for their
support.

A partial support from the grant GA\v CR 202/02/0735 of the Czech
Republic is also acknowledged.
\end{acknowledgements}

\end{document}